\documentclass[vecphys]{svmult}     
\usepackage{cite}
\usepackage{graphicx}        
\usepackage{epsfig}          
\begin{document}

\title*{Experiments on multidimensional solitons}
\author{
J.~Brand\inst{1} \and
L.~D.~Carr\inst{2} \and B.P.~Anderson\inst{3} }

\institute{
Centre of Theoretical Chemistry and Physics,
Institute of Fundamental Sciences,
Massey University Auckland, New Zealand.
\texttt{J.Brand@massey.ac.nz}
\and
Department of Physics, Colorado School of Mines, Golden, CO 80401,
U.S.A. \texttt{lcarr@mines.edu} \and College of Optical Sciences,
University of Arizona, Tucson, AZ 85721, U.S.A.
\texttt{brian.anderson@optics.arizona.edu} }
\maketitle

\section{Dimensional Aspects of Soliton Experiments in Bose-Einstein Condensates}

The experimental work on solitons in BECs during the 
past decade 
has been extremely important in 
two respects.
First, experiments have shown unambiguously that solitons and
related nonlinear waves do exist in BECs. This
was disputed at the time when the first experiments were performed
in
1999~\cite{Burger1999a,Denschlag2000a} and may still be found
surprising given that we are dealing with a quantum many-body system
driven towards strongly non-equilibrium dynamics.  Second, the
experiments have inspired theoretical work in many directions.
A particular example is the work on solitary waves in dimensional
crossovers induced by trapping potentials that restrict the
geometry, as discussed in Chap.~Va.

A ubiquitous feature in all of the experimental realizations of
solitons so far is the three-dimensional nature of experimental
set-ups, which can never be completely neglected in the
interpretation of the results obtained. In this respect almost all
of the experimental work discussed in this book falls into the wider
area of
``multidimensional solitons.''  In order to avoid the duplication of
discussions led in other chapters, we chose to narrow the scope of
this chapter to discuss mainly solitary waves with a genuinely
multidimensional structure.  We therefore restrict the discussion to
experimental studies of the creation and dynamics of vortex rings in
experiments at
JILA~\cite{Anderson2001a} and
Harvard~\cite{Dutton2001a,ginsberg05}.
In particular, we completely omit the vast body of experimental work
on line vortices and the related baby-skyrmions, i.e., vortices with
filled cores~\cite{Matthews1999b,Anderson2000a}), as these will be
covered in Chap.~VI.

We also completely omit the experiments on bright solitons,
and the first planar dark soliton
experiments~\cite{Burger1999a,Denschlag2000a} that were primarily
concerned with soliton observation and propagation in BECs rather
than dynamical instabilities and soliton decay.
Detailed reviews of
these experiments can be found in Chap.~II and Chap.~III,
respectively.
%

\section{Preparation of non-equilibrium BECs}

None of the solitons and nonlinear waves discussed in this chapter
are ground states.
Rather, they are defects in the background density of the BEC. In
order to facilitate the generation of such defects, the condensate
is brought into an unstable or genuinely time-dependent dynamical
state,
such as a planar dark soliton in a single-component condensate. More
stable multi-dimensional solitons can then form spontaneously as the
decay product of an unstable initial state.
%
In particular, vortex rings have been created by decay of moving or
stationary planar dark solitons. The planar solitons themselves have
been created by density engineering, which will be discussed below,
by phase engineering, as discussed in Chap.~III, or by a combination
of both, which may be called \emph{quantum state
engineering}. Two different realizations of quantum state engineering have
been proposed in Refs.~\cite{Williams1999c,carr01b}.

\subsection{Dark Soliton Quantum State Engineering}


The method used at JILA  to create planar dark
solitons~\cite{Ch3}
in BECs is an example of quantum state engineering that manipulates
both the BEC density and
phase~\cite{Anderson2001a}. The technique
relies
on the ability to place a BEC into a superposition of two
spatially overlapped components, where both the relative quantum
phase and the amplitudes of the superposition components
can
be engineered to have a desired variation across the
BEC~\cite{Williams1999c}. For the JILA experiments with $^{87}$Rb, a
BEC has phase-coherent components created from the internal
hyperfine atomic states $|1\rangle \equiv |F=1, m_F=-1\rangle$ and
$|2\rangle \equiv |F=2, m_F=1\rangle$.  The two components are
coupled with a precisely
engineered
two-photon microwave field, enabling the overall superposition to be
manipulated.
To visualize this, let $|\psi\rangle$ be a state vector which
represents the spatial 
and temporal variations of the two-component BEC.  Then the
superposition takes the form
\[ |\psi\rangle = c_1 e^{i
\phi_1}\cdot|1\rangle  + c_2 e^{i \phi_2}\cdot|2\rangle ,\]
%
where, $c_1$, $c_2$, $\phi_1$, and $\phi_2$ are real scalars which
depend on time and space.  Then $|c_j|^2$ is the number density of
the $j^{\mathrm{th}}$ component and $\phi_1-\phi_2$ is the relative
phase.  Two-component phase and density manipulation requires
control of $c_1$, $c_2$, and $\phi_1-\phi_2$ in order to produce a
final engineered superposition.  Once the desired superposition is
created, the microwave coupling drive is removed.

In the JILA experiment, the two atomic states effectively shared a
potential well in all spatial coordinates.  The two components also
have nearly identical intra- and inter-component scattering lengths.
Thus the sum $|c_1|^2 +|c_2|^2$ remained approximately constant in
space and time during microwave-induced internal state
conversion, i.e., the \emph{total} atomic density of the BEC did not
significantly change, regardless of the spatial structure of the
superposition.
For example, if $\phi_2$ is constant across the BEC,
and $\phi_1-\phi_2=0$ for $z<0$ and $\pi$ for $z>0$, where $z$
labels the vertical spatial direction, there will be a $\pi$ phase
jump across the $z=0$ plane of the part of the BEC made up of
$|1\rangle$ atoms.  This phase jump corresponds to the phase jump
across a horizontal dark soliton nodal plane.

With these general concepts, planar dark solitons were created in a
nearly spherical potential with a mean trapping frequency of $\sim
7.7$ Hz and at temperatures of $T\approx23$ nK, or
$T/T_c\approx0.8$. With an initial BEC made entirely of $\sim 10^6$
atoms in state $|2\rangle$, a two-photon microwave field coupled the
two atomic states 
inducing transitions $|2\rangle \to |1\rangle$
while a laser-induced
AC Stark shift altered the detuning of the microwave field from the
atomic resonance.  The beam was modulated across the BEC such that
the coherent atomic transitions on opposite sides of the BEC were
driven out of phase, resulting in a relative phase of $\pi$ between
the two halves of the component-$|1\rangle$ BEC. 
Additionally, transfer of atoms from $|2\rangle$
to $|1\rangle$ at the center of the BEC atom cloud was suppressed.
Engineered superpositions of
a dark soliton state in component
$|1\rangle$ and uniform-phase states of component $|2\rangle$ were
thus created. More precisely,
a \emph{filled} dark soliton was created in component $|1\rangle$,
with atoms in component $|2\rangle$ occupying the dark soliton nodal
plane in
$|1\rangle$, since the total BEC atomic density $|c_1|^2 +|c_2|^2$
remained approximately constant in time and
space.
The dark soliton remained dynamically stable as long as the
component-$|2\rangle$ ``filling'' remained intact. To study planar
dark soliton dynamical instabilities, including observations of
soliton decay into vortex rings, the $|2\rangle$ atoms were first
removed with a short blast of a laser beam tuned to an electronic
transition of the $|2\rangle$ atoms. The subsequent observations
will be described in
Sec.~\ref{sec:decay}.

The JILA state-engineering technique represents a general method for
creating topological states, as described by Williams and
Holland~\cite{Williams1999c}. For example, the two-component
engineering technique was also used to create vortices in
BECs~\cite{Matthews1999b}, and prior experimental results may be
interpreted in terms of the creation of a vertical stack of filled
horizontal dark solitons in a single
BEC~\cite{Matthews1999a}.
Additional details and references regarding this technique, and
results of other experiments may be found in the experimental
portion of
Chap. IX of this volume, and in the references cited above.

\subsection{Density Engineering by Slow Light}
\label{sec:DensEng}

The method used in the lab of Lene Hau at Harvard to create
nonlinear
waves~\cite{Dutton2001a,ginsberg05} may be described as
density engineering of the BEC. There are two stages in the
process. First, a density depletion of the size scale of a couple of
micrometers, or several healing lengths, is created on a
microsecond
timescale. In the next stage, the condensate reacts on the
millisecond time scale by developing shock waves which shed soliton
wave fronts.
Here we describe the creation of the initial density depletions.

\begin{figure}[hbt]
\center
\includegraphics[width=10cm]{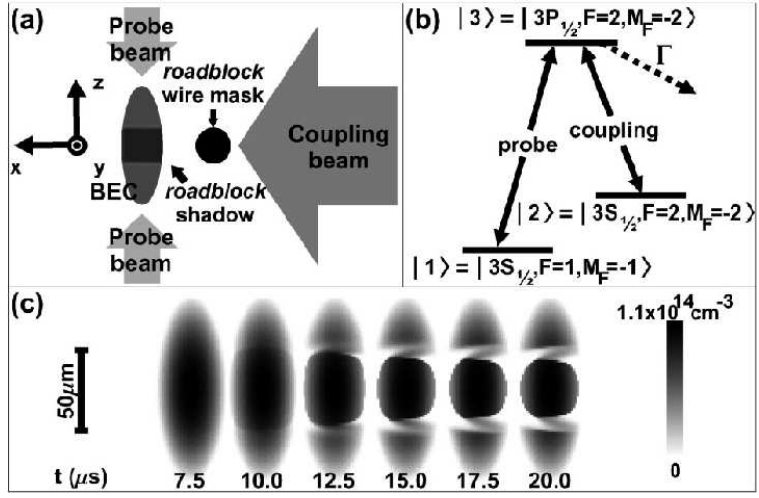}
\caption{Creation of defects by density engineering, from
  \cite{ginsberg05}.  (a) Schematic 
of experimental set up for
  slow-light propagation via EIT and double light
  roadblock. (b) Energy level diagram of Na for EIT and
  slow-light. (c) Two-dimensional simulation of slow-light propagation
  on the double road block showing the density of atoms in state
  $|1\rangle$.} \label{fig:ExpSchem}
\end{figure}

The scheme for creating density depletions is based on the method of
ultra-slow light pulse propagation by {\em electromagnetically
induced transparency}
(EIT)~\cite{boller91prl-EIT,field91prl-EIT}
and an extension thereof termed {\em roadblock} for
light~\cite{Hau1999a,Liu2001a}. The experiments work with a rather 
large BEC of 1.5 to a few million
$^{23}$Na atoms. The effect of EIT makes use of the level structure
of $^{23}$Na and involves mainly the three states
labeled $|1\rangle$, $|2\rangle$, and $|3\rangle$, in
Fig.~\ref{fig:ExpSchem}b, where $|1\rangle$ is the ground state. A
laser pulse tuned exactly on the $|1\rangle-|3\rangle$ transition
(labeled ``probe'') would not be able to propagate in a BEC of atoms
in the $|1\rangle$ state. The medium would absorb all the photons
and appear opaque. If, however,
the levels $|2\rangle$ and $|3\rangle$ are coupled by another laser
field, the $|1\rangle-|3\rangle$ transition line splits into two
lines and photons cannot be absorbed at the original transition
frequency due to interference of the transition amplitudes. This
effect is called EIT because the 
medium becomes transparent at the original transition frequency.
Another consequence of this scenario is that the index of refraction
$n$ varies strongly around the EIT frequency. This, in turn, leads
to a greatly reduced group velocity which is indirectly proportional
to the frequency gradient of $n$. In this way the speed of light was
reduced to a group velocity of 17~m/s~\cite{Hau1999a,dutton04}.

The reduction of the group velocity of light leads to a compression
of light pulses in space 
by the same factor. In this way, a few microsecond long laser
pulse of a kilometer length in vacuum is compressed to about
50$\mu$m, at which point it is completely contained in the BEC. When
the coupling laser is turned off abruptly at this point, light
propagation stops completely and the energy and phase of the probe
beam are stored in the condensate in the form of atoms in the
$|2\rangle$ state. By turning the coupling beam back on after a time
window of the order of
microseconds, the light pulse can be re-released
coherently~\cite{Liu2001a}.

The idea behind the ``light roadblock'' is to transfer the above
described technique of stopping light from the temporal to the
spatial domain. The coupling beam does not illuminate the entire BEC
cloud
because part of it is shaded by means of a razor
blade~\cite{Dutton2001a} or a wire mask~\cite{ginsberg05}.
Figure~\ref{fig:ExpSchem}(a) shows a schematic of the experimental
set-up of a double roadblock with a wire mask and two slow-light
probe beams from Ref.~\cite{ginsberg05}.

\begin{figure}[hbt]
\center
\includegraphics[clip,width=6cm,viewport=255 555 540 1021]{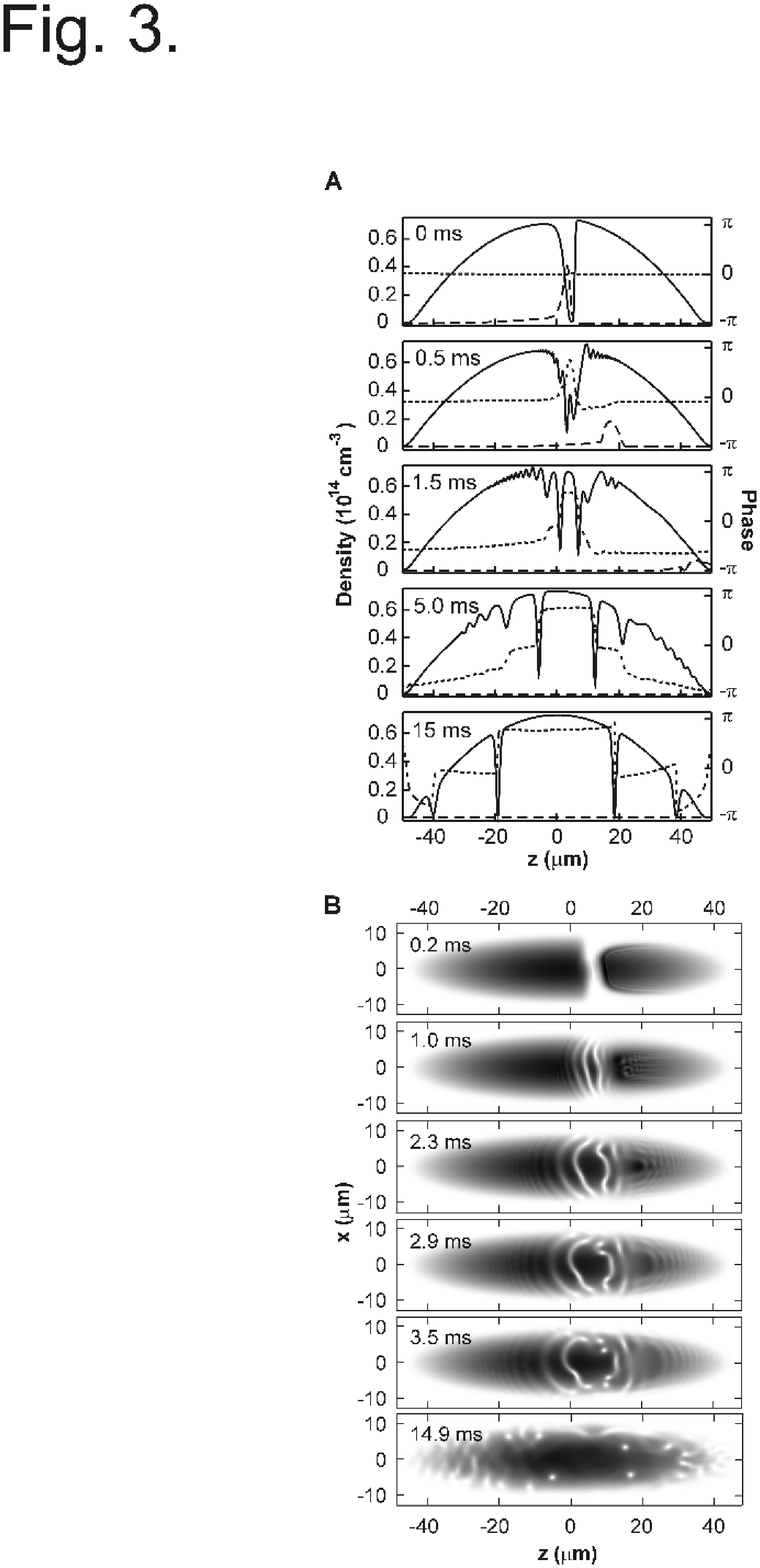}
\includegraphics[clip,width=6cm,viewport=255 102 532 546]{duttonfig3.eps}
\caption{Simulations of slow-light and BEC dynamics.
  (A)
  One-dimensional simulation of the coupled dynamics of
  light and matter-wave fields showing dynamics after the stopping of
  a light pulse on a single roadblock. The solid and dashed lines show
  the densities of $|1\rangle$ atoms and $|2\rangle$
  atoms, respectively. The phase
  of the condensate in $|1\rangle$ is shown by the dotted line. Atoms in
  state $|2\rangle$ are clearly seen to leave the condensate without
  affecting the ensuing dynamics of the state $|1\rangle$ BEC. The evolution
  from density defects into dark solitons is clearly seen from the
  phase profile of the BEC. (B) Two-dimensional simulation of the
  state $|1\rangle$ BEC
  dynamics after creation of a defect. Shown are
  grey-scale plots of the condensate density.
  The
  cigar-like shape of the BEC induced by the trapping potential and the
  development and propagation of dark soliton fronts (white lines) and
  vortices (white dots) are clearly visible.
  From \cite{Dutton2001a}.
  Reproduced with permission from AAAS. } \label{fig:SloDefect}
\end{figure}

As the slow light pulse reaches the edge of the shadowed region of
the BEC, the group velocity is reduced
further, which in turn compresses the pulse further. At the size scale
of a few microns, the stopped light transfers atoms from state
$|1\rangle$ into state $|2\rangle$. Since the atoms in state
$|2\rangle$ carry a photon-induced recoil momentum, they are ejected
out of the condensate and leave within 1~ms without contributing
much to the dynamics of the BEC of atoms in state $|1\rangle$.
Figure~\ref{fig:SloDefect} illustrates this point clearly in
simulations of the dynamics in a coupled GPE model. The evolution of
the density of state $|1\rangle$ atoms during the slow-light
propagation in the case of a double road block is shown in
Fig.~\ref{fig:ExpSchem}(c).

\section{Decay Dynamics and Formation of Multidimensional Solitons}
\label{sec:decay}

\subsection{Quantum Shock Wave Dynamics and Soliton Shedding}
\label{sec:ShockW}

\begin{figure}[hbt]
\parbox{0.58\textwidth}{
\includegraphics[width=0.58\textwidth]{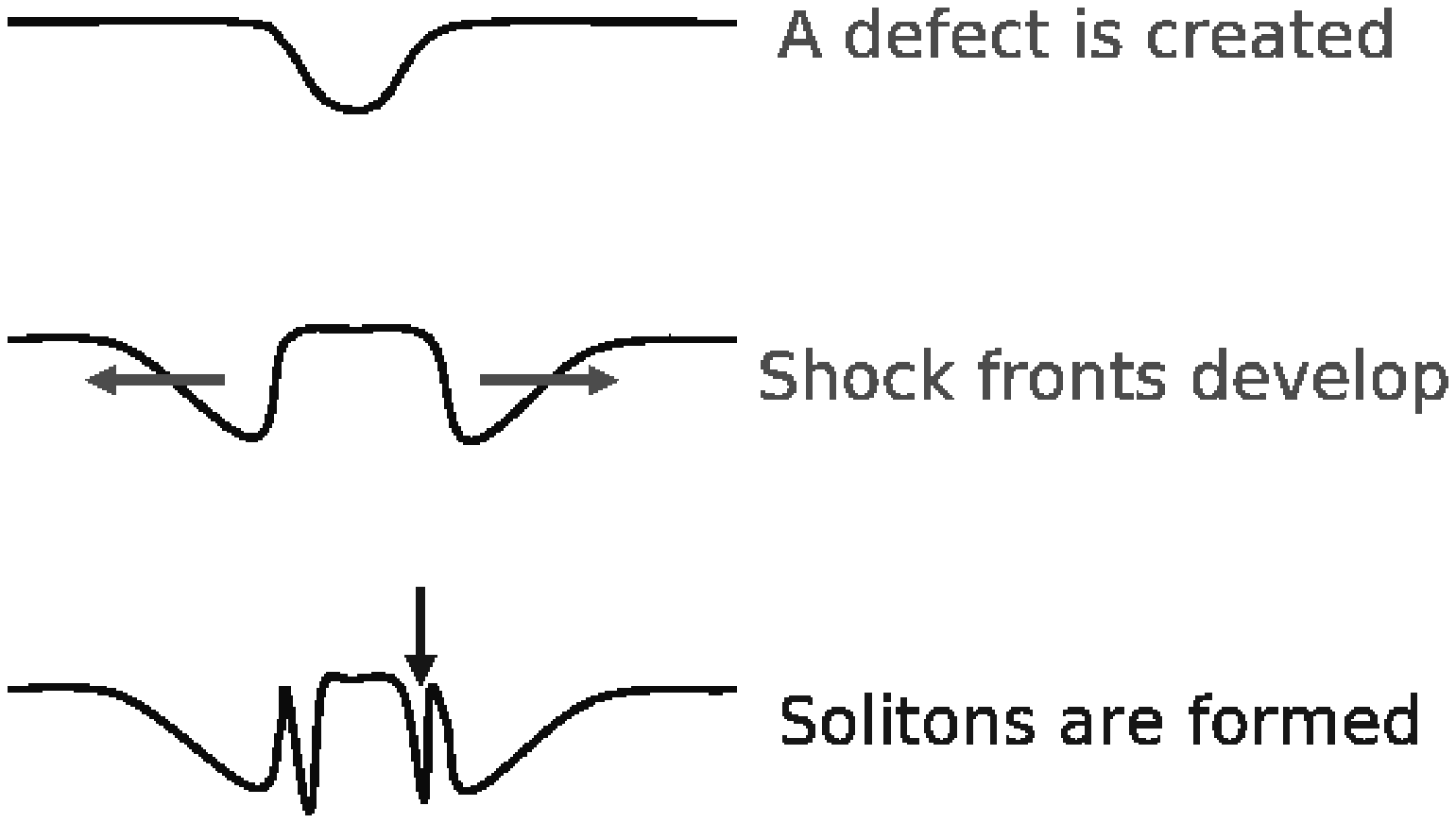}\\[3cm]

\caption{
Top: Schematic of shock-wave formation and soliton shedding
  after creation of a density void.    Right: Experimental data after
  expansion showing wave front dynamics following the creation of a
  single density defect. The slice images
  clearly show planar band soliton fronts at initial stages that
  undergo
  a snake instability and decay into vortex rings. The times cited refer
  to in-trap evolution before the expansion phase. From \cite{Dutton2001a}.
  Reproduced with permission from AAAS.}
\label{fig:DuttonData}}
\hfill
\parbox{0.4\textwidth}{
\includegraphics[draft=false,clip,angle=0,width=0.4\textwidth,bb=265
30 515 779]{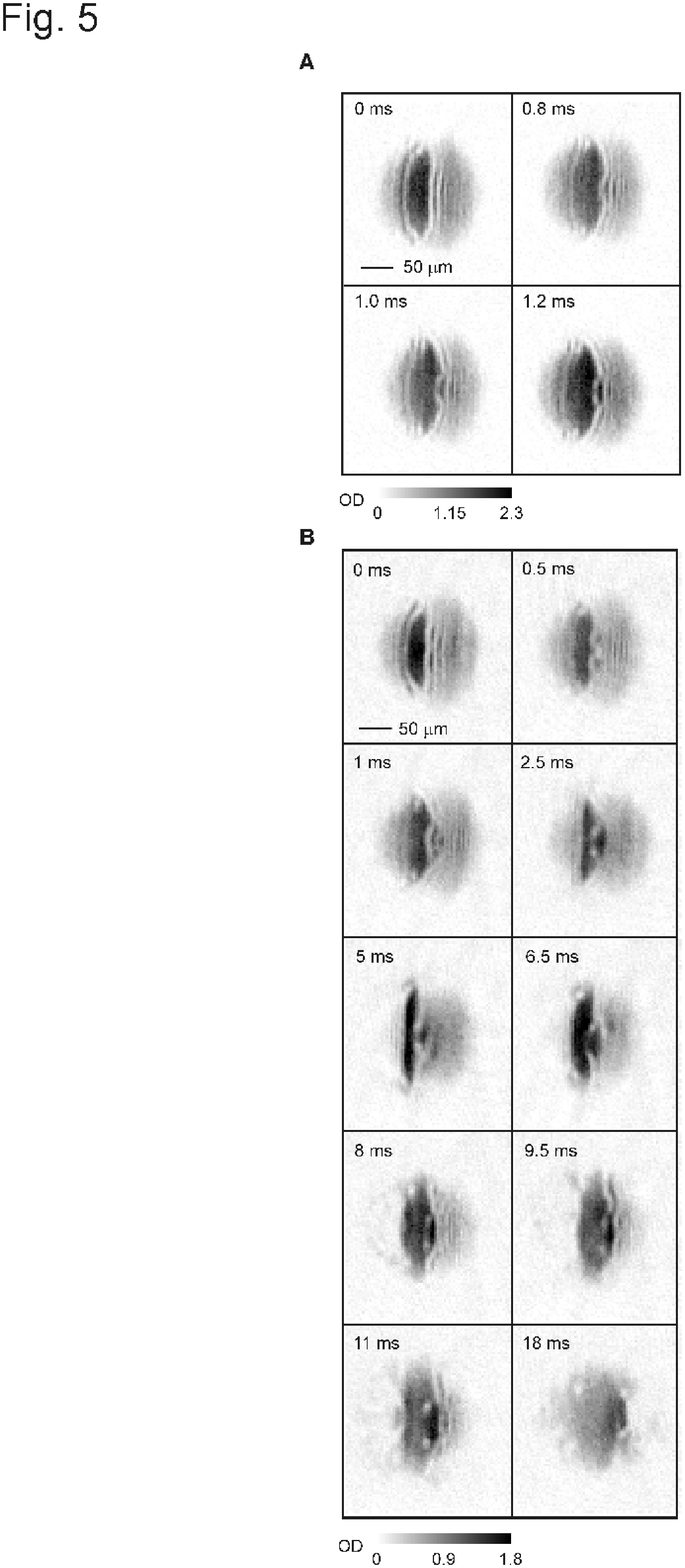}}
\end{figure}

Density engineering as performed in the Harvard
experiments~\cite{Dutton2001a,ginsberg05} and described in
Sec.~\ref{sec:DensEng} primarily creates localized density voids of
the shape of a narrow disk in a slowly-varying background that is
determined by the external trapping potential. Simulations of the
ensuing dynamics are shown in Fig.~\ref{fig:SloDefect}.  The
condensate reacts to the creation of the defect and to the missing
mean-field repulsion from the removed atoms by
rushing in to fill the void. This way, a wave of density depletion
emanates from the void at roughly the local speed of sound.
However, due to the variation of the speed of sound over the density
profile, the wave form steepens at the back to form a shock front.
The variation of the
speed of sound over the density profile is a general nonlinear sound wave
effect, that will become important whenever the density variation is
comparable to the density itself.
Due to quantum pressure in the superfluid hydrodynamics of the BEC,
shock fronts of a size scale smaller than the condensate healing
length are not permitted. Instead, dark solitons in the form of
planar and modulated wave fronts are
shed, as seen in the
simulations of Fig.~\ref{fig:SloDefect} and the schematic of
Fig.~\ref{fig:DuttonData}.

During the propagation of the shock wave, a train of solitons is
generated in its wake. The solitons shed first are the
deepest and equal in amplitude to the original density
wave; then shallower solitons follow. Due to the inverse 
relationship between the depth of the soliton and its speed, the
deep solitons
which were created earlier move more slowly than the ones created at
later times. The deeper solitons also carry a larger amount of
excitation energy, which is taken away from the shock wave. During
its propagation the shock wave will therefore
lose depth and energy and eventually disperse into fast-moving
shallow solitons and sound waves.  This superfluid version of a
shock wave is also called 
a {\em quantum shock wave}.

For theoretical investigations of shock waves in BECs see
Refs.~\cite{zak03:_shock,perez-garcia:220403,kamchatnov:063605,damski04pra}.
The dynamics of shock-wave formation and soliton shedding as described
above could not be observed {\it in situ} due to limits in the optical
resolution for in-trap imaging. However, the resulting trains of dark
soliton fronts are clearly seen in the experimental data shown in
Fig.~\ref{fig:DuttonData}.

Detailed theoretical studies and experimental observations of shock
wave formation and dynamics have been reported recently by Hoefer {\it
  et al.}  \cite{hoefer:023623}. The decay dynamics of the shock
front was described theoretically in terms of a {\em dispersive shock
  wave}.  The shock waves studied in this work 
form from positive density deformations (bright density waves) after
the pulsed application of a strongly repulsive, focussed laser beam to
a BEC rather than from a collapsing cavity  as in the Harvard experiments.
Thus the details of the formation and subsequent dynamics of the shock
wave are different from the situation described above.

\subsection{Snake Instability and Vortex Ring Generation}

The
decay of planar dark soliton fronts and the subsequent formation of 
vortex rings were seen in two quite different experiments
done in the labs of JILA and Harvard and both published in 2001.
The mechanism of decay is a
dynamical instability of planar dark solitons, a  three-dimensional BEC
equivalent of the snake instability of optical dark
solitons~\cite{mamaev96,tikhonenko1996ovs}.
In the following we discuss the observations from both experiments.

\subsubsection{3.2.1 Harvard Experiments}

We discussed above how shock waves caused by collapsing voids have
lead to the creation of dark soliton wavefronts in the Harvard
experiments~\cite{Dutton2001a,ginsberg05}. The dynamics of these
wave fronts and their decay into vortex rings has been probed
experimentally. The size scale of solitons and vortex cores is of
the order of the healing length,
in this case about 1 $\mu$m. In order to increase the size of the
density features to be more easily imaged, the condensate is
suddenly released from the trap and expands due to mean-field
pressure while it falls in the gravitational field of the earth.
After a typical delay time of about 15 to
20 ms, a single slice of the cloud is selectively imaged by
near-resonant absorption after illumination by a pump laser through
a slit mask.

The initial expansion image of Fig.~\ref{fig:DuttonData} clearly shows
a train of soliton wave fronts to the right of the dark density
feature. During the subsequent time evolution the deepest of the
soliton wave fronts undergoes
the snake instability while the shallower solitons to the right
appear to be more stable. The snaking of the deep soliton front
eventually leads to the creation of a vortex ring. The signature of
a vortex ring in the experimental images are pairs of white dots.
The dots show the depleted core region of the vortex ring
intersecting the plane of imaging twice. The slice imaging technique
was used to tomographically scan the expanded cloud
and explore the three-dimensional structure of the observed defects.
The experimental data was found consistent with the assumption that
the
snake dynamics and vortex generation obey
cylindrical symmetry for most of the
process, except for late stages of the time evolution as seen in the
bottom panels of Fig.~\ref{fig:DuttonData}. This observation
suggests that the
snake dynamics are seeded by geometrical preconditioning,
i.e.,
bending of soliton fronts due to the geometry of the created
defects, trapping potential, etc., rather than being seeded by
thermal or quantum fluctuations of the field.

Extensive simulations of the dynamics of the BEC
both in the trap and during the
expansion phase have revealed that nonlinear wave dynamics still
take place during the
latter~\cite{ginsberg05}.
Therefore, theoretical modeling is
essential for the understanding and interpretation of the
experimental results.

\subsubsection{3.2.2 JILA Experiments}


Using two-component state engineering to create dark solitons in
BECs, the JILA group also observed dark solitons decay into vortex
rings~\cite{Anderson2001a}. The underlying mechanism of decay is
presumed to be again the snake instability, although  in the JILA
experiment the actual ``snaking'' was not
resolved, as was the case in the Harvard
experiment.

As described in Section 2.1, planar dark solitons at JILA were
formed in two-component condensates, with atoms of component
$|2\rangle$ filling the nodal plane of a planar dark soliton in
component $|1\rangle$. The soliton state and the filling could be
separately distinguished with in-trap state-selective phase-contrast
imaging, i.e., \emph{without} releasing the BEC from the trap.
This
confirmed that the soliton's density notch bisected the BEC. In an
orthogonal coordinate system with axes labeled by $x,y$ and $z$,
with $z$ along the vertical, the soliton was always created with a
normal vector in the $y-z$ plane. Although the soliton orientation
was not further controlled, images of filled solitons along the
$x$-direction revealed the orientation of each soliton created.
Simultaneous images taken along the $y$ direction also showed the
soliton plane on the occasions when it was horizontal. This
non-destructive two-axis imaging technique was used to identify the
initial orientation of the soliton nodal planes of the BECs for
correlation with subsequent images of soliton decay.

\begin{figure}[t]
\center
\includegraphics[width=12cm]{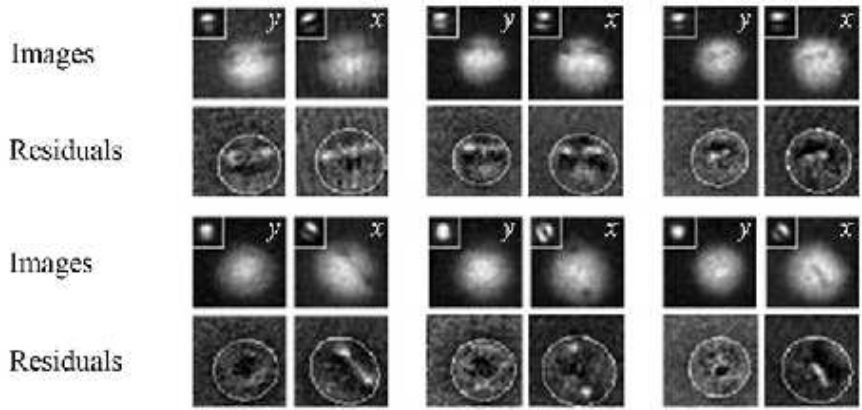}
\caption{Images of filled dark solitons (insets) and subsequent
expansion images with density dips that correspond to atomic fluid
displaced by vortex rings.  Each image set consists of two expansion
images simultaneously acquired along the $x$ and $y$ directions, and
residuals obtained after subtracting a Thomas-Fermi (TF) fit of each
expansion image. Outlines of the TF surface are shown as white
ellipses in the residuals; the lighter regions within correspond to
fluid depletion in the BEC.  Note that in these images of
condensates, the light-dark areas are reversed from the images of
the Harvard condensates. Each full-size image represents an area of
220 $\mu$m $\times$ 220 $\mu$m.  The figure is similar to Figure 4
of
Ref.~\cite{Anderson2001a}.}
\label{fig_vortexrings}
\end{figure}

To investigate and study dynamical instabilities of
\emph{single}-component dark solitons, the component-$|2\rangle$
filling was removed with a
100 ms blast of laser light resonant with an electronic transition
of the $|2\rangle$ atoms, which left behind a dark soliton in
component $|1\rangle$.  The trapped BEC was then held for a variable
time, typically less than 100 ms, before release from the confining
potential and absorption imaging after 56 ms of expansion. Dark
solitons were not observed in expansion images, even though the
corresponding in-trap images showed that filled dark solitons were
indeed created. Instead, pairs of density dips in the expanded
images were observed. For $x$-axis images, two density dips were
often seen to lie along a line that approximately matched the
orientation of the corresponding filled dark soliton image for that
same BEC, taken just before the soliton filling was removed. For
example, if an in-trap image along $x$ showed that the dark soliton
was aligned in a horizontal plane, the two density dips would be
seen in the expansion image along an imaginary horizontal line.  If
a soliton did happen to lie in such a horizontal plane, the soliton
nodal plane was then also seen with in-trap images acquired along
the $y$ direction. The corresponding $y$-direction expansion images
in these cases also showed two density dips.
Example images are
shown in Fig. \ref{fig_vortexrings}.

The pairs of density dips in the expansion images
indicate the
presence of vortex rings in the BECs, the predicted decay products
of dynamical instabilities of dark solitons in three-dimensional
BECs~\cite{Feder2000a}. Similar to a line vortex
(see Chap. VI of this volume), and described earlier in this
chapter, a vortex ring in a BEC is characterized by the absence of
atoms in a toroidal region within the BEC, with quantized superfluid
flow around the fluid-free toroid. An $x$-direction column density
profile of a BEC with a vortex ring then appears as a smooth atom
cloud with two ``holes'' separated by the vortex ring diameter, with
a faint line connecting the holes.
However, if the faint line can not be seen then single-directional
imaging does not easily distinguish between
single vortex rings and
a pair of vortex lines. Therefore, two-directional imaging was used
to identify the presence of ring-like structures, as two
density-dips were seen in the expansion images from two orthogonal
directions for the cases where the in-trap images showed the initial
dark soliton nodes to be horizontal.

The experimental results indicated that solitons decayed to vortex
rings by the end of the
100 ms removal of the $|2\rangle$ atoms. The
observations were consistent with numerical simulations, also
described in
Ref.~\cite{Anderson2001a}.  However, while numerical results
predicted the presence of up to three vortex rings, the experimental
results revealed no more than a single vortex ring per image.
Nevertheless,
the experimental results confirmed the theoretical expectations of
soliton decay, and demonstrated that multi-dimensional vortex rings
were
stable topological soliton structures that could be created and
observed in BECs.


\section{Interacting Dark Solitons and Hybrid Structures}

In the Harvard experiment reported in Ref.~\cite{ginsberg05},
collisions of soliton fronts and vortex rings were probed. In order
to facilitate the interaction between nonlinear waves, density
engineering was used to generate density voids in two different
locations within the same BEC simultaneously. A schematic of the
experimental set up is shown in
Fig.~\ref{fig:ExpSchem}(a) and a simulation of the defect creation
stage is shown in Fig.~\ref{fig:ExpSchem}(c). The mechanism of
soliton shedding from shock waves as described in
Sec.~\ref{sec:ShockW} leads to dark soliton fronts emanating from
two centers. In the central part of the BEC this leads to the
head-on collision of trains of dark soliton fronts.

\begin{figure}[t]
\center
\includegraphics[width=\textwidth]{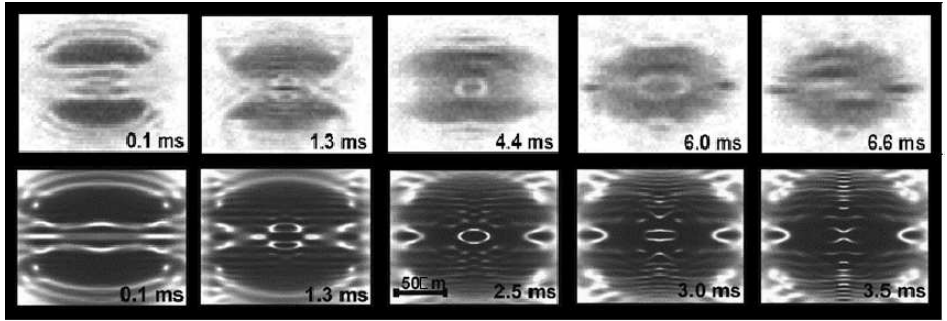}
\caption{Experiment (top row) and simulation (bottom row) of
  nonlinear-wave collisions from
  Ref.~\cite{ginsberg05}.  The
  apparent agreement is a strong indicator for the validity of the
  Gross-Pitaevskii approach to describe the non-equilibrium dynamics
  of BECs.}
\label{fig:HarExptSim}
\end{figure}

Figure \ref{fig:HarExptSim} shows the experimental images of the
condensate density (top row) and the results of a corresponding
Gross-Pitaevskii simulation (bottom row). The images show unexpected
intermediate structures in the form of low-density shells around a
high-density core of oblate to spherical geometry. These structures
are reminiscent of nonlinear Bessel-function-type stationary
solutions of the nonlinear Schr\"odinger equation
\cite{theocharis:120403,theocharis:023609,carr:043613,ginsberg05}
as discussed in
Chap.~Va. In the experiment, the shell structures are clearly of
transient nature and may be closely related to the transient shell
structures seen in simulations of head-on collisions of vortex
rings~\cite{komineas:110401}.

\begin{figure}[hbt]
\center
\includegraphics[angle=270,width=\textwidth]{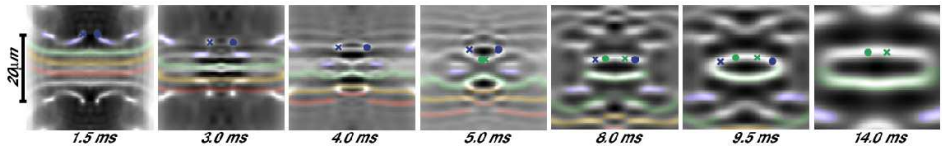}
\caption{Simulation of nonlinear-wave evolution during free expansion
  phase, from
  Ref.~\cite{ginsberg05}. Phase singularities from vortex
rings in the cylindrically symmetric BEC are marked in the upper halves
of the frames by crosses or dots. The sequence shows the intricate
dynamics that eventually leads to the formation of hybrid low-density
shell structures composed of vortex rings and dark soliton fronts.}
\label{fig:HarEvSim}
\end{figure}

Further insight into the structure of the observed shell structures
and their formation can be extracted from the GPE simulations of the
experiment as shown in Figs.~\ref{fig:HarExptSim} and
\ref{fig:HarEvSim}. These simulations revealed that the shell
structures are not present while the BEC is in the trap but rather
are formed during the expansion phase. The intricate dynamics that
leads to the formation of shell structures involves
decay of planar dark soliton fronts via the snake instability into
vortex rings, repeated collisions of vortex rings with soliton
fronts, the propelling and bending of soliton fronts by the
inhomogeneous velocity fields associated with vortex rings, and
finally the reconnection processes between soliton wave fronts and
annihilation of vortex ring pairs. The simulated time evolution
during the expansion phase is shown in Fig.~\ref{fig:HarEvSim}.

\section{Conclusions}

Clearly the experimental investigation of multidimensional solitons
has just begun.
In particular, as three-dimensional wave-guide structures on atom
chips or in the form of toroidal traps become available, it will
become possible and
indeed, quite important to study the propagation modes of solitary
waves in these geometries. Furthermore, as the theoretical work
reviewed in Chap. Va suggests, there is much to be expected from the
experimental investigation of topological solitons and other
nonlinear waves in spinor condensates.
We note that the spontaneous
formation of topological defects such as spin vortices has been
observed very recently in an experiment at
Berkeley~\cite{sadler06:spinBEC} by quenching a spinor BEC through a
ferromagnetic phase transition.  

The authors are indebted to
Zac Dutton for a careful reading of the manuscript.
JB thanks Lene Hau, Naomi Ginsberg, and Sean Garner for many useful
discussions and hospitality during his visits at Harvard University.
LDC gratefully acknowledges the support of the National Science
Foundation. BPA acknowledges support from the National Science
Foundation and the Army Research Office.

\bibliographystyle{mio_bec_book}

\begin{thebibliography}{10}

\bibitem{Burger1999a}
S.~Burger, K.~Bongs, S.~Dettmer, W.~Ertmer, K.~Sengstock, A.~Sanpera, G.~V.
  Shlyapnikov, and M.~Lewenstein.
\newblock {\em Phys. Rev. Lett.} {{\bf 83}}, 5198 (1999).
\bibitem{Denschlag2000a}
J.~Denschlag, J.~E. Simsarian, D.~L. Feder, Charles~W. Clark, L.~A. Collins,
  J.~Cubizolles, L.~Deng, E.~W. Hagley, K.~Helmerson, W.~P. Reinhardt, S.~L.
  Rolston, B.~I. Schneider, and W.~D. Phillips.
\newblock {\em Science} {{\bf 287}}, 97 (2000).
\bibitem{Anderson2001a}
B.~P. Anderson, P.~C. Haljan, C.~A. Regal, D.~L. Feder, L.~A. Collins, C.~W.
  Clark, and E.~A. Cornell.
\newblock {\em Phys. Rev. Lett.} {{\bf 86}}, 2926 (2001).
\bibitem{Dutton2001a}
Z. Dutton, M. Budde, C. Slowe, and L.~V. Hau.
\newblock {\em Science} {{\bf 293}}, 663 (2001).
\bibitem{ginsberg05}
N.~S. Ginsberg, J.~Brand, and L.~V. Hau.
\newblock {\em Phys. Rev. Lett.} {{\bf 94}}, 040403 (2005).
\bibitem{Matthews1999b}
M.~R. Matthews, B.~P. Anderson, P.~C. Haljan, D.~S. Hall, C.~E. Wieman, and
  E.~A. Cornell.
\newblock {\em Phys. Rev. Lett.} {{\bf 83}}, 2498 (1999).
\bibitem{Anderson2000a}
B.~P. Anderson, P.~C. Haljan, C.~E. Wieman, and E.~A. Cornell.
\newblock {\em Phys. Rev. Lett.} {{\bf 85}}, 2857 (2000).
\bibitem{Williams1999c}
J.~E. Williams and M.~J. Holland.
\newblock {\em Nature} {{\bf 401}}, 568 (1999).
\bibitem{carr01b}
L.~D. Carr, J.~Brand, S.~Burger, and A.~Sanpera.
\newblock {\em Phys. Rev. A} {{\bf 63}}, 051601(R) (2001).
\bibitem{Ch3}
See Chapter III of this volume for theoretical and experimental reviews of dark
  solitons in BECs.
\bibitem{Matthews1999a}
M.~R. Matthews, B.~P. Anderson, P.~C. Haljan, D.~S. Hall, M.~J. Holland, J.~E.
  Williams, C.~E. Wieman, and E.~A. Cornell.
\newblock {\em Phys. Rev. Lett.} {{\bf 83}}, 3358 (1999).
\bibitem{boller91prl-EIT}
K.-J. Boller, A.~Imamolu, and S.~E. Harris.
\newblock {\em Phys. Rev. Lett.} {{\bf 66}}, 2593--2596 (1991).
\bibitem{field91prl-EIT}
J.~E. Field, K.~H. Hahn, and S.~E. Harris.
\newblock {\em Phys. Rev. Lett.} {{\bf 67}}, 3062--3065 (1991).
\bibitem{Hau1999a}
L.~V. Hau, S.~E. Harris, Z. Dutton, and C.~H. Behroozi.
\newblock {\em Nature} {{\bf 397}}, 594 (1999).
\bibitem{Liu2001a}
C. Liu, Z. Dutton, C.~H. Behroozi, and L.~V. Hau.
\newblock {\em Nature} {{\bf 409}}, 490 (2001).
\bibitem{dutton04}
Z.~Dutton, N.~S. Ginsberg, C.~Slowe, and L.~V. Hau.
\newblock {\em Europhysics News} {{\bf 35}} (2004).
\bibitem{zak03:_shock}
M. Zak and I. Kulikov.
\newblock {\em Phys. Lett. A} {{\bf 307}}, 99--106 (2003).
\bibitem{perez-garcia:220403}
V.~M. Perez-Garcia, V.~V. Konotop, and V.~A. Brazhnyi.
\newblock {\em Phys. Rev. Lett.} {{\bf 92}}, 220403 (2004).
\bibitem{kamchatnov:063605}
A.~M. Kamchatnov, A.~Gammal, and R.~A. Kraenkel.
\newblock {\em Phys. Rev. A} {{\bf 69}}, 063605 (2004).
\bibitem{damski04pra}
B. Damski.
\newblock {\em Phys. Rev. A} {{\bf 69}}, 043610 (2004).
\bibitem{hoefer:023623}
M.~A. Hoefer, M.~J. Ablowitz, I.~Coddington, E.~A. Cornell, P.~Engels, and
  V.~Schweikhard.
\newblock {\em Phys. Rev. A} {{\bf 74}}, 023623 (2006).
\bibitem{mamaev96}
A.~V. Mamaev, M.~Saffman, and A.~A. Zozulya.
\newblock {\em Phys. Rev. Lett.} {{\bf 76}}, 2262 (1996).
\newblock and A. V. Mamaev {\it et al.}, Phys. Rev. A {\bf 54}, 870 (1996).
\bibitem{tikhonenko1996ovs}
V.~Tikhonenko, J.~Christou, B.~Luther-Davies, and Y.S. Kivshar.
\newblock {\em Opt. Lett} {{\bf 21}}, 1129--1131 (1996).
\bibitem{Feder2000a}
D.~L. Feder, M.~S. Pindzola, L.~A. Collins, B.~I. Schneider, and C.~W. Clark.
\newblock {\em Phys. Rev. A} {{\bf 62}}, 053606 (2000).
\bibitem{theocharis:120403}
G.~Theocharis, D.~J. Frantzeskakis, P.~G. Kevrekidis, B.~A. Malomed, and
  Yuri~S. Kivshar.
\newblock {\em Phys. Rev. Lett.} {{\bf 90}}, 120403 (2003).
\bibitem{theocharis:023609}
G.~Theocharis, P.~Schmelcher, M.~K. Oberthaler, P.~G. Kevrekidis, and D.~J.
  Frantzeskakis.
\newblock {\em Phys. Rev. A} {{\bf 72}}, 023609 (2005).
\bibitem{carr:043613}
L.~D. Carr and Charles~W. Clark.
\newblock {\em Phys. Rev. A} {{\bf 74}}, 043613 (2006).
\bibitem{komineas:110401}
S. Komineas and J. Brand.
\newblock {\em Phys. Rev. Lett.} {{\bf 95}}, 110401 (2005).
\bibitem{sadler06:spinBEC}
L.~E. Sadler, J.~M. Higbie, S.~R. Leslie, M.~Vengalattore, and D.~M.
  Stamper-Kurn.
\newblock {\em Nature} {{\bf 443}}, 312--315 (2006).
\end{thebibliography}


\end{document}